\documentclass[fleqn,usenatbib]{mnras}

\usepackage{newtxtext,newtxmath}
\usepackage[T1]{fontenc}

\DeclareRobustCommand{\VAN}[3]{#2}
\let\VANthebibliography\thebibliography
\def\thebibliography{\DeclareRobustCommand{\VAN}[3]{##3}\VANthebibliography}

\usepackage{graphicx}	% Including figure files
\usepackage{amsmath}	% Advanced maths commands
\usepackage{csvsimple}  % Added by Lex, used to import csv tables

\title[Transit timings for four hot Jupiters]{A search for transit timing variations in the transiting hot Jupiter systems HIP~65, NGTS-6, NGTS-10 and WASP-173}

\author[Griffiths et al.]{A.\ W.\ Griffiths,$^{1}$%\thanks{E-mail: a.w.griffiths@keele.ac.uk}
J.\ Southworth,$^{1}$%\thanks{Email: taylorsouthworth@gmail.com}
L.\ Alegre,$^{2,3}$
F.\ Amadio,$^{4}$
M.\ I.\ Andersen,$^{5}$
A.\ J.\ Barker,$^{6}$
M.\ Basilicata,$^{7}$ \newauthor
M.\ Bonavita,$^{4}$
V.\ Bozza,$^{8, 9}$
M.\ J.\ Burgdorf,$^{10}$
R.\ E.\ Cannon,$^{2}$
G.\ Columba,$^{11}$
M.\ Dominik,$^{12}$
A.\ Donaldson,$^{2}$ \newauthor
R.\ Figuera Jaimes,$^{13,14,15,12}$
T.\ C.\ Hinse,$^{16}$
M.\ Hundertmark,$^{17}$
U.\ G.\ J{\o}rgensen,$^{4}$
E.\ Khalouei,$^{18}$ \newauthor 
P.\ Longa-Pe{\~n}a,$^{19}$
L.\ Mancini,$^{20,7,21}$
F.\ Manni,$^{7}$
B.\ Murphy,$^{2}$
N.\ Peixinho,$^{22}$
M.\ Rabus,$^{23}$
S.\ Rahvar,$^{24}$
\newauthor 
H.\ Rendell-Bhatti,$^{2}$
P.\ Rota,$^{8}$
A.\ Ro{\.z}ek,$^{2}$
S.\ Sajadian,$^{24}$
J.\ Skottfelt,$^{25}$
C.\ Snodgrass$^{2}$ and
J.\ Tregloan-Reed$^{19}$
\\
$^{1}$ Astrophysics Group, Keele University, Staffordshire, ST5 5BG, UK [E-mail: a.w.griffiths@keele.ac.uk (AWG), taylorsouthworth@gmail.com (JS)]\\
$^{2}$ Institute for Astronomy, University of Edinburgh, Royal Observatory, Edinburgh EH9 3HJ, UK \\
$^{3}$ Centre for Astrophysics Research, Department of Physics, Astronomy and Mathematics, University of Hertfordshire, College Lane, Hatfield AL10 9AB, UK \\
$^{4}$ Centre for ExoLife Sciences, Niels Bohr Institute, Jagtvej 155, 2200 Copenhagen, Denmark \\
$^{5}$ Cosmic Dawn Centre (DAWN), Niels Bohr Institute, Jagtvej 155, 2200 Copenhagen, Denmark \\
$^{6}$ Department of Applied Mathematics, School of Mathematics, University of Leeds, Leeds, LS2 9JT, UK \\
$^{7}$ INAF -– Osservatorio Astrofisico di Torino, Via Osservatorio 20, 10025, Pino Torinese, Italy \\
$^{8}$ Dipartimento di Fisica ``E.R. Caianiello'', Università di Salerno, Via Giovanni Paolo II 132, 84084, Fisciano, Italy \\
$^{9}$ Istituto Nazionale di Fisica Nucleare, Sezione di Napoli, Napoli, Italy \\
$^{10}$ Universit{\"a}t Hamburg, Faculty of Mathematics, Informatics and Natural Sciences, Department of Earth Sciences, Meteorological Institute, Bundesstra\ss{}e \\ ~~~~~~~~~~ 55, 20146 Hamburg, Germany \\
$^{11}$ Alma Mater Studiorum - University of Bologna, Dipartimento di Fisica e Astronomia ``Augusto Righi'', Via Gobetti 93/2, 40129 Bologna, Italy \\
$^{12}$ Centre for Exoplanet Science, SUPA, School of Physics \& Astronomy, University of St Andrews, North Haugh, St Andrews KY16 9SS, UK \\
$^{13}$ Instituto de Astronomia y Ciencias Planetarias, Universidad de Atacama, Copayapu 485, Copiapo, Chile \\
$^{14}$ Millennium Institute of Astrophysics MAS, Nuncio Monsenor Sotero Sanz 100, Of.\ 104, Providencia, Santiago, Chile \\
$^{15}$ Instituto de Astrof\'{\i}sica, Facultad de F\'{\i}sica, Pontificia Universidad Cat\'olica de Chile, Av.\ Vicu\~na Mackenna 4860, 7820436, Macul, Santiago, Chile \\
$^{16}$ University of Southern Denmark, Department of Physics, Chemistry and Pharmacy, SDU-Galaxy, Campusvej 55, 5230, Odense M, Denmark \\
$^{17}$ Astronomisches Rechen-Institut, Zentrum f{\"u}r Astronomie der Universit{\"a}t Heidelberg (ZAH), 69120 Heidelberg, Germany \\
$^{18}$ Astronomy Research Center, Research Institute of Basic Sciences, Seoul National University, 1 Gwanak-ro, Gwanak-gu, Seoul, 08826, Korea \\
$^{19}$ Centro de Astronom{\'{\i}}a, Universidad de Antofagasta, Av.\ Angamos 601, Antofagasta, Chile \\
$^{20}$ Dipartimento di Fisica, Universit\`a degli Studi di Roma Tor Vergata, via della Ricerca Scientifica 1, 00133, Roma, Italy \\
$^{21}$ Max Planck Institute for Astronomy, K\"onigstuhl 17, 69117, Heidelberg, Germany \\
$^{22}$ Instituto de Astrof\'{\i}sica e Ci\^encias do Espa\c{c}o, Departamento de F\'{\i}sica, Universidade de Coimbra, 3040-004, Coimbra, Portugal \\
$^{23}$ Departamento de Matem\'atica y Física Aplicadas, Facultad de Ingeniería, Universidad Católica de la Santísima Concepción, Alonso de Rivera 2850, \\ ~~~~~~~~~~ Concepci\'on, Chile \\
$^{24}$ Perimeter Institute for Theoretical Physics, 31 Caroline St N, Waterloo, ON N2L 2Y5, Canada \\
$^{25}$ Centre for Electronic Imaging, Department of Physical Sciences, The Open University, Milton Keynes, MK7 6AA, UK \\
}

\date{Accepted XXX. Received YYY; in original form ZZZ}

\pubyear{\the\year{}}

\begin{document}
\label{firstpage}
\pagerange{\pageref{firstpage}--\pageref{lastpage}}
\maketitle

\begin{abstract}
Hot Jupiters are Jupiter-mass planets with orbital periods of less than ten days. Their short orbital separations make tidal dissipation within the stellar host especially efficient, potentially leading to a measurable evolution of the orbit. One possible manifestation of this is orbital decay, which presents itself observationally through variations in the orbital period and thus times of transit. Here we select four promising exoplanetary systems for detecting this effect: HIP\,65, NGTS-6, NGTS-10 and WASP-173. We present 33 new transit light curves taken with the 1.54\,m Danish Telescope, and analyse these alongside photometric data from the Transiting Exoplanet Survey Satellite and transit timing data from the literature. We construct two ephemeris models for each target: a linear ephemeris and a shrinking orbital period due to tidal decay. The linear ephemeris is preferred for three of the four models -- the highest significance for the quadratic ephemeris is over 3$\sigma$ for WASP-173. We compare these results to theoretical predictions for tidal dissipation of gravity waves in radiation zones, and find that wave breaking is predicted only in WASP-173, making rapid decay plausible in this system but unclear in the other three. The sensitivity of transit timings to orbital decay depends on the square of the time interval covered by available observations, so our results establish a useful baseline against which future measurements can be compared. NGTS-6 and NGTS-10 are important objects for future study as they are in the first field to be observed by the upcoming PLATO mission.

%This is a simple template for authors to write new MNRAS papers.
%The abstract should briefly describe the aims, methods, and main results of the paper.
%It should be a single paragraph not more than 250 words (200 words for Letters).
%No references should appear in the abstract.
\end{abstract}


%%%%%%%%%%%%%%%%%%%%%%%%%%%%%%%%%%%%%%%%%%%%%%%%%%

%%%%%%%%%%%%%%%%% BODY OF PAPER %%%%%%%%%%%%%%%%%%

\section{Introduction}

Hot Jupiters, classified based on their large masses and short orbital periods, are a relatively rare type of exoplanet within the underlying population \citep{2010Sci...330..653H,2012ApJS..201...15H,mayor2011harpssearchsouthernextrasolar}. Despite their scarcity, they make up a significant fraction of the known exoplanets \citep{Gaudi_2002,2012ApJ...753..160W}. This is a consequence of their relatively large sizes and small semimajor axes making them the easiest transiting planets to detect. This proximity to the host star raises tides on both bodies, causing tidal bulges \citep{Murray_Dermott_2000}.

For a planet in an asynchronous orbit -- one in which the planet's spin period is not equal to the orbital period -- the motion of the tidal bulges is opposed by viscous friction. This dissipation of energy transfers angular momentum from the planetary orbit to the stellar spin. In the typical case for a hot Jupiter, where its orbital period is shorter than the host star's spin period, tidal effects spin up the star and shrink the orbit \citep{1980A&A....92..167H,1981A&A....99..126H}. The eventual outcome of sustained orbital shrinkage is the planet slowly spiralling into the star in a process called tidal decay \citep{1973ApJ...180..307C}. It is worth noting that there are various contributions to tidal dissipation \citep{1996ApJ...470.1187R}, but arguably the largest is that of the wave-like dynamical tide \citep{2020MNRAS.498.2270B}.

%\textbf{See also these papers: 2009ApJ...692L...9L, 2008ApJ...678.1396J, 2009ApJ...698.1357J (and others, probably).}

Tidal decay in exoplanets is not easily detectable. There is currently no confirmed way to measure the spin or internal density distribution of exoplanets, which would otherwise give information on the tidal mechanisms and angular momentum transfer within the system. However, some claims have been made for the observation of oblateness brought on by tides (e.g.\ \citealt{2024ApJ...976L..14L,2025ApJ...981L...7P,2025ApJ...983..157Z}). The timescales on which tides act are also unknown and it has been argued that the circularization of transiting exoplanets may not be entirely due to tides \citep{2009ApJ...692L...9L}. 

Presently, signs of tidal decay have only been detected through transit-timing variations (TTVs; \citealt{2018haex.bookE...7A}), which show variations in the times of mid-transit and a decrease in the orbital period of the planet over time. The associated change in period is measurable from transit light curves and allows for the determination of the modified tidal quality factor, $Q_{\star}'$, providing an insight into the rate of energy dissipation within the star due to tides. $Q_{\star}'$ is defined as follows:
\begin{equation}
Q_{\star}' = \frac{3 Q_{\star}}{2 k_{2}}
\end{equation}
where $k_{2}$ is the Love number \citep{1911spge.book.....L} and $Q_{\star}$ is the tidal quality factor -- linked to the dissipation of tidal energy through the equation \citep{1966Icar....5..375G}:
\begin{equation}
    Q_{\star}=\frac{1}{2\pi E_0}\oint\bigg(-\frac{dE}{dt}\bigg)dt
\end{equation}
where $E_0$ is the maximum energy stored in a tide, $-dE/dt$ is the dissipation rate and the integral is over one tidal cycle. While a negative period derivative can be a response to the tidal evolution of the system, this effect can be caused by other phenomena such as the light-time effect, in which a wide-orbiting third body causes the star-planet pair to orbit the system’s barycentre \citep{1952ApJ...116..211I}, and  apsidal motion, arising only in orbits with non-zero eccentricity. Both phenomena are periodic so can be distinguished from tidal decay over a large enough timescale.

%\textbf{JKT: note that there have just been a few claims of the detection of oblateness using JWST light curves -- check this out. I also suggest arXiv2410.05408 and arXiv.2410:07977.}

WASP-12 b, a 1.5\,M$_{\rm Jup}$ hot Jupiter orbiting an F-type main-sequence star \citep{Hebb_2009}, was until recently the sole confirmed case of tidal decay in exoplanets \citep{Maciejewski+16aa,2020ApJ...888L...5Y,2021AJ....161...72T}. Since the confirmation of the system's shrinking orbital period, the decay rate has been refined multiple times \citep{2022MNRAS.512.3113B,2022AJ....163..281T,2022AJ....163..175W,2023TJAA....4...10A,2023JAVSO..51..243N,2024A&A...685A..63A,2024A&A...686A..84L,2024ApJS..270...14W}. A recent study by \cite{2025A&A...694A.233B} has detected orbital decay within WASP-43, making it the second confirmed hot Jupiter system. Despite the contemporary exclusivity of these systems, there are many candidates for tidal decay, with few having established TTVs. 

%\textbf{JKT: I think there are other objects now. Check WASP-43 (arXiv this week).}
%\textbf{LG: Yup. I've changed this now to include the new one.}

One such candidate is WASP-4 b \citep{2008ApJ...675L.113W}, for which TTVs were first detected by \cite{2019AJ....157..217B}. These authors suggested the apparent period change could be caused by tidal decay, apsidal precession or gravitational perturbation from another body. \cite{2019MNRAS.490.4230S} refined the rate of period shrinkage and ruled out stellar activity and the Applegate mechanism \citep{1992ApJ...385..621A}. \cite{2020ApJ...893L..29B} and \cite{2023Univ....9..506H} investigated the possibility of a line-of-sight acceleration, attributed to a distant perturbing body. With their new transit timing data, \cite{2025MNRAS.541..714B} established that the discrepancies would be best explained by tidal orbital decay. However, the TTVs of the WASP-4 system are now believed to be associated with the light-time effect \citep{2025arXiv251005229W}.

%\textbf{Check out also WASP-4: 2020ApJ...893L..29B, 2019AJ....157..217B, 2019MNRAS.490.4230S}

\subsection{Target selection}

The magnitude of detectable orbital decay within hot Jupiter planetary systems can be quantified using the transit time shift equations presented by \cite{2014MNRAS.440.1470B} and \cite{2018AcA....68..371M}:
\begin{equation} \label{eq:tshift}
T_{\rm shift} = -\frac{27}{4} \frac{\pi}{Q_{\star}'} \biggl(\frac{M_{\rm p}}{M_{\star}}\biggl)\biggl(\frac{R_{\star}}{a}\biggl)^{5} \biggl(\frac{1}{P}\biggl) (10\rm yr)^2 
\end{equation} 
where $T_{\rm shift}$ is the shift in transit midpoints over ten years, $M_{\star}$ and $M_{\rm p}$ are the respective stellar and planetary masses, $R_{\star}$ is the stellar radius, $a$ is the orbital semimajor axis and $P$ is the orbital period. Here, $Q_{\star}'$ is set to a canonical value of $10^6$ \citep{2014ARA&A..52..171O}. Whilst requiring some preliminary spectroscopic parameters, this equation allows systems with the most favourable predicted time shifts to be selected for analysis. 

We used this equation to select targets with the largest $T_{\rm shift}$. We then restricted the sample based on sky position to include only those objects with good observability for the available telescope time. A nominal lower limit of 10\,s was specified for $T_{\rm shift}$, but objects close to this limit were ultimately not observed due to the limited availability of telescope time. The first result of this project was an analysis of HATS-18 \citep{2022MNRAS.515.3212S}; see also a study of KELT-16 by \citet{2022MNRAS.509.1447M}.

%\textbf{JKT: the target selection was more random than this, but we can say we observed objects which were visible in the right part of the year for the telescope time we had. I suggest you also reproduce the Maciejewski equation here. Did it initially come from Birkby et al (2014MNRAS.440.1470B)?}
%\textbf{LG: I've added the equation and a sentence to clarify. You're right, I've found it in Birkby et al. too! Added the reference.}

\subsubsection{HIP\,65\,Ab}
\label{sec:111}

Our first target is HIP\,65, a wide binary star system hosting a K-dwarf (HIP\,65\,A) and an M-dwarf (HIP\,65\,B). The hot-Jupiter planet HIP\,65\,Ab orbits the K-dwarf with an ultra-short period  of 0.981\,d \citep{Nielsen_2020,2021AJ....162..176P}. HIP\,65\,A and HIP\,65\,B are separated by 3.95\,arcsec on the sky \citep{Nielsen_2020}, corresponding to a projected separation of 269\,au. These authors highlighted the planet's large impact parameter and grazing eclipse, making radius estimations unreliable. They stated also that HIP\,65\,Ab is one of very few massive planets that reside within twice their Roche limit, lying on the border of the Neptune desert \citep{Szabó_2011}. However, its exact relation to the Roche limit is uncertain due to the grazing eclipses. \citet{Nielsen_2020} also report HIP\,65\,A's rotation period as $13.2^{+1.9}_{-1.4}$\,d from rotational modulation in light curves, and suggest tidal spin-up due to the discrepancy between the gyrochronological ($0.32^{+0.10}_{-0.06}$\,Gyr) and isochronal ($4.1^{+4.3}_{-2.8}$\,Gyr) ages of the star. However, the isochronal ages of K-dwarfs are unreliable so this is not in itself evidence of tidal evolution (\citealt{Maxted++15aa2}; but see also \citealt{2021ApJ...919..138T}).

Photometric data on HIP\,65\,Ab from the Transiting Exoplanet Survey Satellite (TESS; \citealt{2015JATIS...1a4003R}) have previously been examined for eclipse depth variations \citep{2022sas..conf....7G,2024AJ....167....1W}, with the purpose of tracking its atmospheric activity. No eclipse depth variations were found after fixing the impact parameter; as such, the grazing eclipse was likely responsible for the initial variations.

Using equation\,\ref{eq:tshift}, we find HIP\,65\,Ab to be a promising candidate for detecting tidal decay, having an estimated $T_{\rm shift}$ of $-23.7 \pm 2.2$ seconds in 10 years (s in 10\,yr). In spite of the promise of this system, there have been two previous orbital decay searches of HIP\,65\,Ab with limited success. \cite{2024PSJ.....5..163A} found a period increase of $22.3 \pm 9.6 \rm \,ms\,yr^{-1}$ using 106 transit times spanning five years. \cite{2024A&A...692A..35M} found the decay rate to be $(-0.1 \pm 3.1)\times 10^{-10}$ days per orbital cycle. Both studies were limited by the data available at the time, with the majority of their transit midpoints originating from TESS. Here, we use new precise transit photometry alongside TESS to search for variations in transit timing.

%\textbf{JKT: add more detail. How many transits did Adams have? etc.}
%\textbf{LG: I'll add more detail to this and point out the recent Maciejewski et al study of HIP65.}
%\textbf{Here you should discuss the chances of seeing orbital decay for HIP 65. It's one of the better ones, and this can be quantified using equation 2 from Maciejewski et al (2018AcA....68..371M). See also Birkby (2014MNRAS.440.1470B) and Patra+ (2017) etc.}

\subsubsection{NGTS-6\,b}
\label{sec:112}

Our second target is an ultrashort-period hot Jupiter orbiting an old K-dwarf. NGTS-6\,b \citep{2019MNRAS.489.4125V} is the least massive planet in our sample, 1.3\,M$_{\rm Jup}$, and has an orbital period of 0.882\,d. The estimated orbital decay associated with this planet is $T_{\rm shift} = -19.1 \pm 13.7\,\rm s$ in 10\,yr. \cite{2020MNRAS.498.2270B} suggested that the stellar host likely experiences no wave breaking, so rapid decay is not expected. However, photometric observations may provide signs of tidal decay in the event that its internal gravity waves are fully damped. This exoplanet also orbits near Roche contact \citep{2022ApJ...939...91A} similar to the case of HIP\,65\,Ab.

%\textbf{JKT: this sentence is a bit too informal. Please reword.}
%\textbf{LG: Changed to ``Our second target is an ultrashort-period hot Jupiter orbiting an old K-dwarf.''}

\subsubsection{NGTS-10\,b}
\label{sec:113}

NGTS-10\,b is a 2.16\,M$_{\rm Jup}$ planet with an ultrashort period of 0.767 days and had the shortest known period of any hot Jupiter at the time of discovery \citep{2020MNRAS.493..126M}. Its stellar host is an old K-type star of mass 0.70\,M$_{\odot}$, which has a rotation period of 17.3\,d. There have been various theoretical studies of the system, some investigating its atmosphere and climate (e.g.\ \citealt{2023A+A...671A.122H,2024ApJ...975...61K}), and some investigating the effects of tides (e.g.\ \citealt{2020MNRAS.498.2270B,2021MNRAS.506.2247A,2024ApJ...973..128T}). NGTS-10\,b is an interesting object for several research areas, patticularly orbital decay. We estimate a large $T_{\rm shift}$ of $-62.8 \pm 29.9\,\rm s$ in 10\,yr, but as with NGTS-6, no wave breaking or rapid decay is projected \citep{2020MNRAS.498.2270B}.

%\textbf{Mention something about the host star - its mass or spectral type or something.}
%\textbf{LG: Added ``Its stellar host is an old K-type star of mass 0.70\,M$_{\odot}$.''}

\subsubsection{WASP-173\,Ab}
\label{sec:114}

WASP-173\,Ab (KELT-22\,A) is a hot Jupiter orbiting a solar-type star, and is notable for being discovered by two independent consortia simultaneously \citep{2019MNRAS.482.1379H,2019ApJS..240...13L}. Similarly to the HIP\,65 system, its host star has a distant binary companion, this time at 1400\,au. WASP-173\,Ab is an inflated planet, with a mass of 3.7\,M$_{\rm Jup}$, a radius of 1.2\,R$_{\rm Jup}$, a period of 1.39\,d, and an irradiation level significantly more than the threshold for inflation \citep{2011ApJS..197...12D,2019ApJS..240...13L}. Its stellar host is a G-type star with a rotation period of around $8.4\,\rm d$ \citep{2024A&A...690A.379K}.

%\textbf{Is there an errorbar for this age?}
%\textbf{LG: Removed the age thing}

This planet has seen little individual attention since its discovery, mostly being included in tide-related projects with substantial sample sizes (e.g. \citealt{2021ApJ...919..138T}). The linear ephemeris has been refined numerous times in previous large-scale studies (e.g. \citealt{2022ApJS..259...62I,2023ApJS..264...37S,2023ApJS..265....4K}) and analysed for timing variations \citep{2024PSJ.....5..163A,2024A&A...692A..35M}, with many suggesting it is a good candidate for tidal decay (e.g. \citealt{2020AJ....159..150P,2024ApJ...960...50W}). With the time shift equation, we estimate a period decay of $-23.9 \pm 6.6\,\rm ms$ over the next 10 years and contrast it with the increase in orbital period found by \cite{2024PSJ.....5..163A} of $19.3 \pm 11.0 \rm\,ms\,yr^{-1}$, and that of $(1.1 \pm 5.2)\times10^{-10}$ days per cycle by \cite{2024A&A...692A..35M}. We aim to refine these prior studies with the addition of our new transit light curves, whilst also providing data for future work in this area.

%\textbf{JKT: no it isn't. See WASP-86 / KELT-12 (2022Obs...142....1S), and there are others}
%\textbf{LG: I've changed 'unique' to 'notable'}
%\textbf{I also expect you to manage to insert a reference to my HATS-18 paper in here somehow. 2022MNRAS.515.3212S.}

\section{Observations}

\subsection{Danish 1.54-m Telescope}
\label{section21}

\begin{table*} \centering
\caption{\label{tab:obslog} Log of the transit observations obtained for this work. $N_{\rm obs}$ is the number
of observations, $T_{\rm exp}$ is the exposure time, $T_{\rm dead}$ is the mean time between the end of one
exposure and the start of the next, and `Moon illum.' is the fractional illumination of the Moon at the
midpoint of the transit. The aperture radii are target aperture, inner sky and outer sky, respectively.}
\begin{tabular}{llcccccccccccc} \hline
Target & Date of first & Start time & End time  & $N_{\rm obs}$ & $T_{\rm exp}$ (s) & $T_{\rm dead}$ (s) & Filter & Airmass &     Moon     & Aperture \\ % & $N_{\rm poly}$ & Scatter
       & observation   &    (UT)    &   (UT)    &               &               &                &        &         & illumination & (pixels) \\ % &                & (mmag)
\hline
HIP 65   & 2021/09/20 & 04:32 & 07:17 & 230 &    30   & 11 & $I$ & 1.11 $\to$ 1.26            & 0.992 & 11 20 40 \\ % & 0.9 \\ % Bonavita, Rahvar
HIP 65   & 2021/09/22 & 04:06 & 06:31 & 210 &    30   & 11 & $I$ & 1.11 $\to$ 1.19            & 0.983 & 25 35 70 \\ % & 1.1 \\ % Tregloan-Reed, Bonavita
HIP 65   & 2022/09/15 & 05:17 & 06:46 &  75 &    60   & 12 & $R$ & 1.11 $\to$ 1.17            & 0.747 & 20 32 70 \\ % &   ? \\ % Tregloan-Reed, Rajkumar, Romero-Colmenares (telescope technical problem)
HIP 65   & 2022/09/16 & 04:58 & 07:24 & 114 &    60   & 12 & $V$ & 1.11 $\to$ 1.25            & 0.658 & 17 25 45 \\ % & 0.8 \\ % Tregloan-Reed, Burgdorf, Rajkumar, Romero-Colmenares
HIP 65   & 2022/09/21 & 02:51 & 05:14 & 165 &    40   & 12 & $R$ & 1.19 $\to$ 1.11            & 0.212 & 22 32 70 \\ % & 0.9 \\ % Burgdorf, Sajadian
HIP 65   & 2023/07/15 & 08:26 & 10:28 & 101 &    60   & 13 & $I$ & 1.12 $\to$ 1.11 $\to$ 1.14 & 0.054 & 25 40 80 \\ % & 0.6 \\ % Rabus, Sajadian
HIP 65   & 2023/07/17 & 07:21 & 09:48 & 122 &    60   & 13 & $I$ & 1.17 $\to$ 1.11 $\to$ 1.12 & 0.003 & 25 35 80 \\ % & 0.7 \\ % Rabus, Sajadian
HIP 65   & 2023/09/13 & 04:32 & 06:55 & 196 &    30   & 10 & $I$ & 1.12 $\to$ 1.11 $\to$ 1.17 & 0.032 & 20 30 60 \\ % & 0.9 \\ % Rota, Rabus
HIP 65   & 2024/07/18 & 04:31 & 06:50 & 261 &    20   & ~9 & $I$ & 1.67 $\to$ 1.21            & 0.878 & 20 30 60 \\ % & 1.2 \\ % Rabus
HIP 65   & 2024/07/19 & 04:08 & 06:09 & 229 &    20   & 11 & $I$ & 1.78 $\to$ 1.28            & 0.938 & 20 30 60 \\ % & 1.9 \\ % Rabus, Avalos-Vega, Longa-Pena
HIP 65   & 2024/09/04 & 06:21 & 08:15 & 420 &     5   & 10 & $I$ & 1.11 $\to$ 1.25            & 0.014 & 12 24 40 \\ % &   ? \\ % Figuera Jaimes, Khalouei
\hline
NGTS-6   & 2019/08/23 & 07:11 & 10:26 & 235 &    40   & 10 & $R$ & 1.90 $\to$ 1.04            & 0.524 & ~8 18 35 \\ % & 2.7 \\ % Spyratos, Burgdorf, Peixinho ("Carried-out focused due to low counts")
NGTS-6   & 2019/09/07 & 07:53 & 10:01 &  80 &    80   & 11 & $R$ & 1.24 $\to$ 1.03            & 0.630 & 19 28 50 \\ % & 1.7 \\ % Bach-Møller, Spyratos ("Stars are not separated and counts are low")
NGTS-6   & 2019/09/15 & 07:24 & 09:58 & 125 &    60   & 14 & $R$ & 1.23 $\to$ 1.00 $\to$ 1.24 & 0.986 & ~8 20 50 \\ % & 1.4 \\ % Bach-Møller, Antilén, Sajadian ("Stars are separated")
NGTS-6   & 2021/09/25 & 05:11 & 07:15 & 156 &    30   & 15 & $I$ & 1.65 $\to$ 1.14            & 0.832 & ~7 20 40 \\ % & 2.2 \\ % Tregloan-Reed ("Good separation of PSFs with faint close star")
NGTS-6   & 2021/10/02 & 06:12 & 08:29 & 187 &    30   & 13 & $I$ & 1.24 $\to$ 1.01            & 0.202 & ~8 19 40 \\ % & 2.3 \\ % Figuera Jaimes, Hinse (Good separation of PSFs with faint close star")
NGTS-6   & 2022/10/11 & 06:05 & 08:00 &  97 &  35--60 & 14 & $I$ & 1.16 $\to$ 1.01            & 0.977 & ~7 18 35 \\ % & 1.4 \\ % Bonavita, Rabus
NGTS-6   & 2023/10/05 & 05:54 & 07:53 & 148 &    30   & 14 & $I$ & 1.26 $\to$ 1.02            & 0.667 & 13 22 50 \\ % & 2.2 \\ % Longa Pena, Rota
NGTS-6   & 2024/09/05 & 07:48 & 09:59 & 175 &    30   & 13 & $I$ & 1.27 $\to$ 1.02            & 0.047 & ~9 19 40 \\ % & 1.6 \\ % Figuera Jaimes, Khalouei
NGTS-6   & 2024/09/13 & 06:27 & 08:20 & 178 &  30--20 & 13 & $I$ & 1.51 $\to$ 1.10            & 0.711 & ~8 20 40 \\ % & 2.7 \\ % Donaldson
\hline
NGTS-10  & 2022/09/12 & 07:17 & 09:52 & 103 & 100--55 & 12 & $R$ & 1.71 $\to$ 1.07            & 0.951 & ~8 12 50 \\ % & 1.7 \\ % Tregloan-Reed, Rajkumar, Romero-Colmenares
NGTS-10  & 2022/09/22 & 06:39 & 09:08 &  79 &   100   & 14 & $R$ & 1.70 $\to$ 1.08            & 0.128 & ~9 25 40 \\ % & 1.2 \\ % Burgdorf, Sajadian
NGTS-10  & 2023/10/09 & 04:36 & 08:27 & 125 &   100   & 13 & $R$ & 2.46 $\to$ 1.05            & 0.256 & 15 22 40 \\ % & 1.4 \\ % Tregloan-Reed, Figuera Jaimes
NGTS-10  & 2023/10/12 & 06:32 & 08:28 &  63 &   100   & 11 & $R$ & 1.28 $\to$ 1.03            & 0.056 & ~9 12 17 \\ % & 1.2 \\ % Tregloan-Reed, Rahvar
NGTS-10  & 2024/09/18 & 07:10 & 09:32 &  85 &    60   & 13 & $R$ & 1.57 $\to$ 1.06            & 0.999 & ~6 12 50 \\ % & 2.9 \\ % Donaldson, Rendell-Bhatti
NGTS-10  & 2024/09/28 & 06:21 & 09:38 &  13 &    60   & 13 & $R$ & 1.64 $\to$ 1.01            & 0.172 & ~8 25 40 \\ % & 1.4 \\ % Rendell-Bhatti, Sajadian
\hline
WASP-173 & 2021/09/29 & 03:03 & 07:24 & 390 &    20   & 14 & $R$ & 1.03 $\to$ 1.00 $\to$ 1.44 & 0.490 & 21 32 70 \\ % &   ? \\ % Tregloan-Reed, Hinse ("the tails of the two PSFs slightly cross")
WASP-173 & 2022/07/06 & 05:19 & 08:42 & 244 &    30   & 17 & $I$ & 1.59 $\to$ 1.01            & 0.419 & ~7 38 52 \\ % & 1.4 \\ % Crake, Hinse
WASP-173 & 2022/09/05 & 05:14 & 09:12 & 543 &  10--20 & 13 & $I$ & 1.00 $\to$ 1.51            & 0.676 & ~8 30 50 \\ % & 1.7 \\ % Tregloan-Reed, Rajkumar, Romero-Colmenares
WASP-173 & 2023/08/01 & 06:02 & 10:16 & 835 &   5--4  & 13 & $I$ & 1.08 $\to$ 1.00 $\to$ 1.20 & 0.995 & ~8 25 50 \\ % &   ? \\ % Columba, Burgdorf
WASP-173 & 2023/08/26 & 04:49 & 09:15 & 522 &    10   & 17 & $I$ & 1.05 $\to$ 1.00 $\to$ 1.32 & 0.701 & ~8 35 50 \\ % &   ? \\ % Longa Pena, Sajadian
WASP-173 & 2024/07/03 & 04:52 & 09:22 & 317 &    30   & 13 & $I$ & 1.91 $\to$ 1.00            & 0.083 & ~8 30 50 \\ % & 1.6 \\ % Basilicata, Rota
%NOT USED: WASP-173 & 2024/08/14 & 04:39 & 06:43 & 315 &    10   & 13 & $I$ & 1.12 $\to$ 1.00            & 0.757 & ~9 30 50 \\ % &   ? \\ % Cannon, Murphy, Sajadian
WASP-173 & 2024/08/22 & 02:52 & 07:05 & 732 &  10--5  & 13 & $I$ & 1.37 $\to$ 1.00 $\to$ 1.02 & 0.486 & ~9 30 50 \\ % &   ? \\ % Longa-Pena, Chamoun, Carcamo
\hline \end{tabular}
\end{table*}

All new ground-based observations in this work were obtained using the 1.54\,m Danish Telescope at ESO La Silla, Chile, as a side-project of the MiNDSTEp microlensing observations \citep{Dominik+10an}. The Danish Faint Object Spectrograph and Camera (DFOSC) was used in imaging mode, in which form it is equipped with a 2048$\times$2048 pixel CCD camera with a 13.7$^\prime$ field of view sampled at 0.39$^{\prime\prime}$\,px$^{-1}$. The CCD was windowed to decrease the readout time in all cases. The filters used were Johnson $V$, Bessell $R$ and Bessell $I$. Some of the observations were obtained with the telescope moderately defocussed to improve the precision of the observations, following the method set out by \citet{Me+09mn}. An observing log is given in Table\,\ref{tab:obslog}. All observations in the log were included in the timing analysis.

HIP~65 is a bright star ($V=11.1$) and the planet produces a short (47\,min) and shallow (0.6\%) eclipse. We experimented with using different filters and focus levels, in some cases pre-planned and in others as a response to the sky conditions during an observing sequence. The majority of our observations were, in the end, obtained through the $I$ filter and with moderate defocus, and achieved a photometric precision in the region of 1\,mmag per point.

NGTS-6 is rather fainter ($V=14.1$) and has a companion at an angular distance of 5.4$^{\prime\prime}$. The majority of our observations were taken with no or small defocus, and through the $I$ filter to lower the count rate. When the seeing was good we were able to extract light curves of NGTS-6 without contamination by the fainter nearby star.

NGTS-10 is another relatively faint star ($V=14.3$) but without a nearby companion. In all our observing sequences we used an $R$ filter to maximise throughput, and moderate defocussing with exposure times of 60--100\,s. The photometric precisions obtained are in the region of 1.5\,mmag per point.

WASP-173 is a bright star ($V=11.2$) with a fainter companion at 6.0$^{\prime\prime}$. When the observing conditions were good we used short exposure times and operated the telescope in focus to extract light curves of WASP-173 without contamination. In times of poorer seeing we defocussed the telescope and increased the exposure time, with the intention of recording light curves of the combined light of WASP-173 and its companion. This means some of our light curves accurately reflect the properties of WASP-173 itself, whereas some suffer from third light which causes a smaller transit depth.

The data were reduced using the {\sc defot} pipeline \citep{Me+09mn2,Me+14mn}, which in turn uses the {\sc idl}\footnote{{\tt https://www.ittvis.com/idl/}} implementation of the {\sc aper} routine from {\sc daophot} \citep{Stetson87pasp} contained in the NASA {\sc astrolib} library\footnote{{\tt http://idlastro.gsfc.nasa.gov/}} to perform aperture photometry. We constructed master bias and flat-field images but did not apply them to the data because their main effect was to add to the scatter in the light curves without modifying the shape of the transit. A differential-magnitude light curve was generated for each transit observation by constructing an optimal composite comparison star to calculate differential magnitudes against. The composite comparison star was made by iteratively adjusting the weights of individual stars and the coefficients of a low-order polynomial to minimise the scatter in the data outside transit.

The timestamps for the midpoint of each image were taken from the headers of the {\sc fits} files and converted to the BJD$_{\rm TDB}$ timescale using routines from \citet{Eastman++10pasp}. Manual time checks were performed for many transits, in all cases confirming the reliability of the timestamps in the {\sc fits} headers.

\begin{figure*}
    \centering
    \includegraphics[width=1\linewidth]{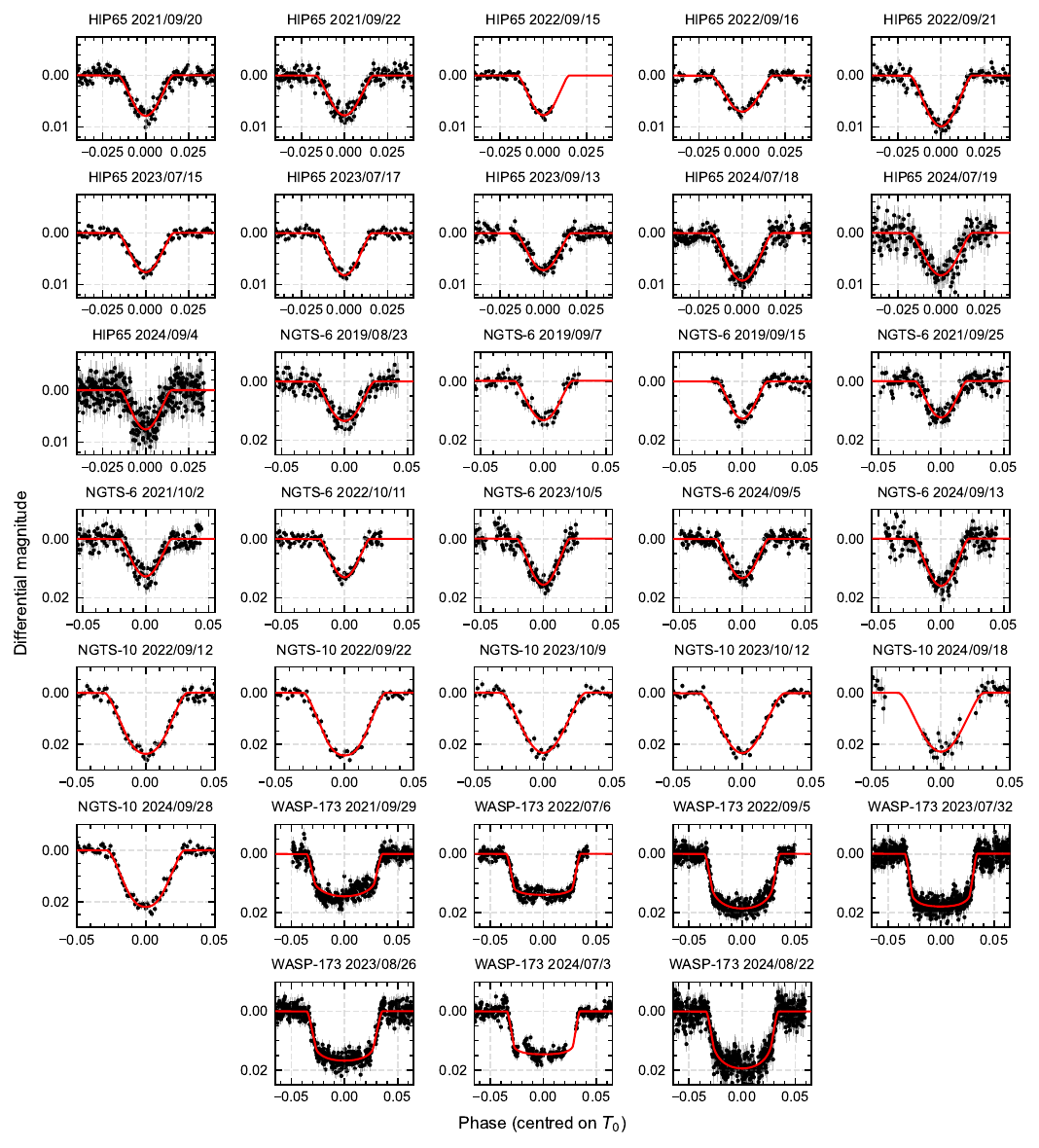}
    \caption{New transit light curves for HIP\,65\,A, NGTS-6, NGTS-10 and WASP-173\,A taken with the Danish Telescope. Data points are displayed in black with their errorbars. Fits are plotted in red. The dates of observation are displayed above each plot.}
    \label{fig:fig1}
\end{figure*}

\subsection{Transiting Exoplanet Survey Satellite}

%\textbf{JKT: put your description of the TESS data here.}

TESS data were extracted from the Mikulski Archive for Space Telescopes (MAST) archive\footnote{https://mast.stsci.edu/portal/Mashup/Clients/Mast/Portal.html} via the Lightkurve package \citep{2018ascl.soft12013L}. We exclusively used 120\,s cadence observations from SPOC, with a standard bitmask. Pre-search Data Conditioning Simple Aperture Photometry (PDCSAP) fluxes \citep{2016SPIE.9913E..3EJ} were used where possible. The first and final 1\% of times in each light curve were discarded to remove any artifacts. We applied the Savitzky-Golay filter \citep{1964AnaCh..36.1627S} to remove low-frequency trends before the light curves were normalized. Fluxes were then converted to magnitude units. Each TESS sector was split into two, where the most central transit midpoint was taken from each half-TESS sector for the light curve and timing analyses. 

HIP\,65\,A has been observed in six TESS sectors (1, 2, 28, 29, 68 and 69) spanning a five-year time interval, with our ground-based transit light curves adding an additional year. WASP-173\,A has been observed in half as many sectors (2, 29 and 69), NGTS-6 was only observed in two sectors (5 and 32) and NGTS-10 has coverage from just one sector (87).

\section{Analysis}

%\textbf{Here describe the \textsc{jktebop} analysis then how you fit the tmin.}
%\textbf{Include a table of all tmin for HIP 65. Reference where each one came from -- those from this paper will be labelled ``This work''.}

\subsection{Light curve analysis}

We used version 43 of the \textsc{jktebop} code \citep{2013A&A...557A.119S} to fit all light curves. Free parameters for our ground-based observations include the sum of fractional radii ($r_{\star} + r_{\rm p}$, where $r_{\rm \star,p}=\frac{R_{\rm \star,p}}{a}$), ratio of the radii ($k=\frac{r_{\rm p}}{r_{\star}}$), inclination ($i$), light scale factor and time of mid-transit ($T_{\rm mid}$). The orbital period ($P$) was also made a free parameter during light curve fitting of TESS data. We applied the power-2 limb darkening law \citep{1997A&A...327..199H}, with coefficients interpolated from tabulations by \cite{2022A&A...664A.128C}. These coefficients were kept fixed. We assume fixed circular orbits for all targets based on the adopted values established in all respective discovery papers, despite some small but non-zero eccentricity results from other sources (e.g.\ \citealt{2023ApJS..265....4K}). All other initial and fixed transit parameters were taken from the respective discovery papers. Uncertainties on fitted parameters were calculated using both Monte Carlo and residual-shift methods in \textsc{jktebop} \citep{2008MNRAS.386.1644S}. The residual-shift method assesses the importance of correlated noise, accounting for it in the error budget. The overall uncertainties on the mid-transit times were picked as the larger of the two multiplied by the corresponding reduced $\chi^2$ value. Fig.\,\ref{fig:fig1} shows all new transit light curves from the Danish Telescope.

\subsection{Literature mid-times}

We combined our transit times for HIP\,65\,Ab with an additional three from the literature. These come from the ExoClock Project \citep{2022ExA....53..547K,2022ApJS..258...40K,2023ApJS..265....4K}, an operation aiming to track the ephemerides of over 1000 transiting exoplanets. The three transit times were taken in succession with the Yves Jongen Telescope in Deep Sky Chile. Three transit times were also pulled from the Exoplanet Transit Database (ETD), a catalogue of transit light curves taken by amateur astronomers \citep{2010NewA...15..297P}.

Eight transit times of NGTS-6\,b are available from the ExoClock Project: four from Yves Jongen at Deep Sky Chile, three from Jean-Pascal Vignes at Deep Sky Chile and one from Ana\"el W\"unsche at El Sauce Observatory. These times span the month of December in the year 2021. We refitted one light curve using the NGTS data published in the planet's discovery paper \citep{2019MNRAS.489.4125V} and three additional times were taken from the ETD. We note that six of the transit midpoints from ExoClock have an epoch in common with an ExoClock transit time. Since all common times are taken with different telescopes, we do not reject any of the data. The same applies to a transit time from the ETD which was obtained at the same epoch as one we observed with the Danish Telescope.

%\textbf{JKT: watch out with the word ``conjunction'' because it means secondary as well as primary eclipses. If you mean transits here, you should say ``transit'' or ``inferior conjunction''.}
%\textbf{LG: Changed to inferior conjunction.}
%We also include the time of inferior conjunction derived in the planet's discovery paper \citep{2019MNRAS.489.4125V} and three additional times from the ETD.

Available timing data for the NGTS-10 system is scarce. The only source in the literature is the discovery paper \citep{2020MNRAS.493..126M}, from which we obtain four transit midpoints by refitting the transit light curves. The resulting times were converted from HJD$_{\rm UTC}$ to BJD$_{\rm TDB}$ following the method described in Section~\ref{section21}. No data are currently available from ExoClock or the ETD.

In the case of WASP-173\,Ab, one transit light curve was taken from the first discovery paper by \cite{2019MNRAS.482.1379H} and refitted. Timing data were given in HJD$_{\rm UTC}$ and so were converted to BJD$_{\rm TDB}$. Three mid-transit times were presented in the second discovery paper \citep{2019ApJS..240...13L}. The ExoClock Project and ETD again both have timing data. We obtained one transit time from each.

\subsection{Transit timing analysis}

\begin{table}
    \centering
    \caption{Parameters and statistical outputs of the linear and quadratic fits for each hot Jupiter.}
    \begin{tabular}{lcc}
        \hline\hline
        \multicolumn{3}{|c|}{HIP\,65\,Ab}\\
        \hline\hline
        Quantity & Linear model & Quadratic model\\
        \hline
        $T_0$ (BJD) & 2459540.547764(42) & 2459540.547697(64)\\
        $P$ (d) & 0.980972186(48) & 0.980972242(63)\\
        $p$ (d) & - & $(1.16 \pm 0.84)\times 10^{-10}$\\
        $\dot{P}$ ($\rm ms\,yr^{-1}$) & - & $7.5 \pm 5.4$\\
        $N_{\rm dof}$ & 27 & 26\\
        $\chi^2$ & 49.9 & 48.0\\
        BIC & 60.0 & 61.5\\
        AIC & 55.9 & 56.0\\
        log $Q_{\rm *,min}'$ & - & $5.16 \pm 0.04$\\
        %$T_{\rm shift}$ (s/10yr) & - & $-23.71 \pm 2.20$\\
        \hline\hline
        \multicolumn{3}{|c|}{NGTS-6\,b}\\
        \hline\hline
        Quantity & Linear model & Quadratic model\\
        \hline
        $T_0$ (BJD) & 2459550.679388(94) & 2459550.67962(14)\\
        $P$ (d) & 0.88205815(11) & 0.88205809(11)\\
        $p$ (d) & - & $(-2.9 \pm 1.2)\times 10^{-10}$\\
        $\dot{P}$ ($\rm ms\,yr^{-1}$) & - & $-20.6 \pm 8.8$\\
        $N_{\rm dof}$ & 23 & 22\\
        $\chi^2$ & 66.0 & 60.5\\
        BIC & 75.7 & 73.4\\
        AIC & 72.0 & 68.5\\
        log $Q_{\rm *,min}'$ & - & $4.28 \pm 0.06$\\
        %$T_{\rm shift}$ (s/10yr) & - & $-26.37 \pm 2.44$\\
        \hline\hline
        \multicolumn{3}{|c|}{NGTS-10\,b}\\
        \hline\hline
        Quantity & Linear model & Quadratic model\\
        \hline
        $T_0$ (BJD) & 2460035.788622(83) & 2460035.78876(10)\\
        $P$ (d) & 0.766893317(50) & 0.76689297(17)\\
        $p$ (d) & - & $(-1.39 \pm 0.66)\times 10^{-10}$\\
        $\dot{P}$ ($\rm ms\,yr^{-1}$) & - & $-11.5 \pm 5.4$\\
        $N_{\rm dof}$ & 10 & 9\\
        $\chi^2$ & 17.6 & 13.0\\
        BIC & 25.0 & 23.0\\
        AIC & 23.6 & 21.0\\
        log $Q_{\rm *,min}'$ & - & $4.93 \pm 0.19$\\
        %$T_{\rm shift}$ (s/10yr) & - & $-30.31 \pm 2.83$\\
        \hline\hline
        \multicolumn{3}{|c|}{WASP-173\,Ab}\\
        \hline\hline
        Quantity & Linear model & Quadratic model\\
        \hline
        $T_0$ (BJD) & 2459486.704505(71) & 2459486.704763(98)\\
        $P$ (d) & 1.38665315(11) & 1.38665272(15)\\
        $p$ (d) & - & $(-7.3 \pm 1.9)\times 10^{-10}$\\
        $\dot{P}$ ($\rm ms\,yr^{-1}$) & - & $-33.2 \pm 8.7$\\
        $N_{\rm dof}$ & 17 & 16\\
        $\chi^2$ & 58.0 & 43.5\\
        BIC & 66.8 & 55.3\\
        AIC & 64.0 & 51.5\\
        log $Q_{\rm *,min}'$ & - & $4.47 \pm 0.12$\\
        %$T_{\rm shift}$ (s/10yr) & - & $-30.31 \pm 2.83$\\
        \hline
    \end{tabular}
    \label{table stats}
\end{table}

We constructed linear and quadratic ephemerides to fit the timing data. The linear fit represents a constant-period orbit in a system with no orbital decay. It is described by 
\begin{equation}
T_{\rm mid} = P E + T_{0}    
\end{equation} 
where $P$ is the constant orbital period, $E$ is the transit epoch and $T_{0}$ is the value $T_{\rm mid}$ at $E = 0$. The quadratic fit represents an orbital decay model, but is also sensitive to the other aforementioned TTV mechanisms. This model is 
\begin{equation}
T_{\rm mid} = p E^2 + P E + T_0
\end{equation}
and invokes a quadratic coefficient $p$, a quantity linked to the rate of period growth $\dot{P}$ by $p = \frac{1}{2}P\dot{P}$. A negative value of $p$ indicates period shrinkage and possible tidal decay. Model parameters for both fits were estimated using an MCMC method with the emcee package \citep{2013ascl.soft03002F}, where the quadratic coefficient $p$ was allowed to be positive or negative. For each model, we let 500 walkers move 20,000 steps, with an arbitrary 1000 step burn-in. We do not fit for periodic trends in this study due to the sparse data coverage.

%priors were estimated with the bootstrap method \citep{10.1214/ss/1177013815} using weighted least-squares fitting, and 

\begin{figure*}
    \centering
    \includegraphics[width=1\linewidth]{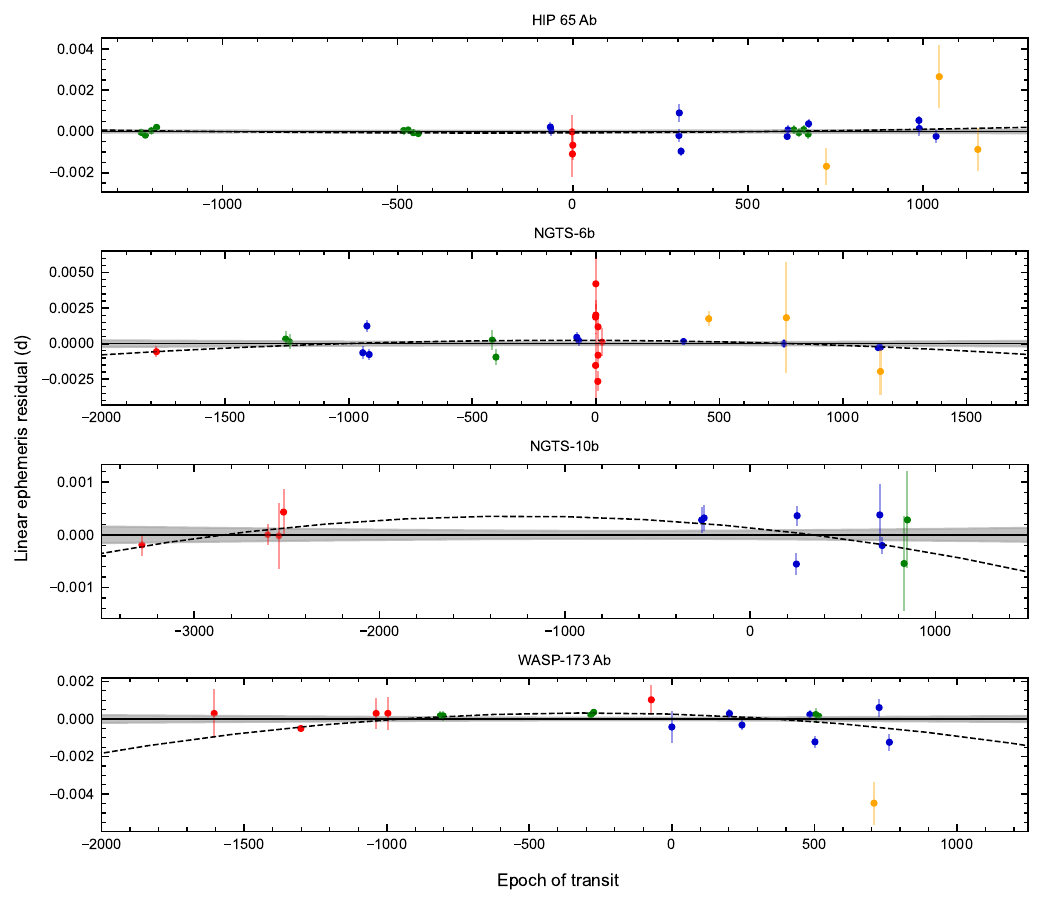}
    \caption{Linear ephemeris residuals plotted against the number of cycles from the median time. The solid line at $y=0$ represents the linear model, where the shaded grey region is its 1$\sigma$ error. The dashed line represents the quadratic model in terms of the linear residuals. Blue points denote new transit times from the Danish Telescope. Green points denote TESS transits. Red points denote transit times from the literature. Orange points denote times from the ETD.}
    \label{fig resid}
\end{figure*}

The difference in strength between the two models was quantified by employing both the Bayesian Information Criterion (BIC; \citealt{10.1214/aos/1176344136}) and Akaike Information Criterion (AIC; \citealt{1100705}). These excel in testing the strength of the fits to the data, whilst also penalising overly-complex higher-order polynomials. The AIC and BIC were computed with 
\begin{equation}
{\rm AIC} = \chi^2 + 2 (n + 1)
\end{equation}
and
\begin{equation}
{\rm BIC} = \chi^2 + \ln (N) (n + 1)
\end{equation}
for each model, where $n$ is the polynomial degree, $N$ is the number of data points and $\chi^2$ was calculated from each model using
\begin{equation}
\chi^2 = \sum_{i=1}^N \Biggl(\frac{T_{{\rm obs,}i} - T_{{\rm calc,}i}}{\sigma _{{\rm obs,}i}}\Biggl)^2
\end{equation}
similarly to equation~2 in \citet{2020AJ....159..150P}. A list of all statistical quantities from the fits is displayed in \autoref{table stats}, and linear residuals are plotted against the best fit in Fig.\,\ref{fig resid}. We compared the two models by subtracting the quadratic BIC and AIC from their linear counterparts. Thus, a significant positive $\Delta$BIC and $\Delta$AIC should imply the presence of TTVs. The range of values of $\Delta$BIC and $\Delta$AIC that indicate a significant preference for one model over another is hard to define, so we required $\Delta{\rm BIC} \ge 10$ and $\Delta{\rm AIC} \ge 10$ to accept the quadratic model in preference to the linear one (e.g. \citealt{Kass01061995}).

%\textbf{JKT: question: why do we get negative Pdot for three systems, whereas previous studies have found no or positive Pdot? Is this just small-number statistics in the available data?}
%\textbf{LG: I've mentioned this briefly in the WASP-173 discussion section.}

\subsection{Tidal quality factor}

We determined a lower bound on the modified tidal quality factor as an indication of the efficiency of tidal dissipation within the star, and its influence over the planet's orbit. Based on results from \cite{2014MNRAS.440.1470B}, \cite{2018AcA....68..371M} and others, and the formulation by \cite{1966Icar....5..375G}, we use the following equation for the modified tidal quality factor: 
\begin{equation}
Q_{\star}' = -\biggl(\frac{27 \pi}{2}\biggl) \biggl(\frac{M_{\rm p}}{M_{\star}}\biggl) \biggl(\frac{R_{\star}}{a}\biggl)^{5} \dot{P}^{-1}
\end{equation}
where $\dot{P}$ is the rate of change of the orbital period. All other parameters have their former meanings. The lower limit on $Q_{\star}'$ for each stellar host was obtained by implementing the 3$\sigma$ lower uncertainty of $\dot{P}$ into the above equation, given the relationship $Q_{\star}' \propto \dot{P}^{-1}$. Limits on $Q_{\star}'$ for the four objects are shown in \autoref{table stats}. Note that this tidal quality factor represents only the stellar tidal dissipation and is independent of any dissipation within the planet.

\subsection{Theoretical predictions}

Our lower bound constraints on $Q'_\star$ for each of the four systems can be compared to theoretical predictions. The most efficient tidal mechanism in each case is predicted to be dissipation of internal gravity waves (part of the dynamical tide response) in their radiative cores, which are launched and propagate inwards from the radiative/convective boundary. If these waves are fully damped (by whatever mechanism) we can readily predict the resulting tidal dissipation and hence $Q'_\star$ (e.g.~using Eq.~41 in \citealt{2020MNRAS.498.2270B}). This regime is likely to be relevant if the planetary mass is large enough to cause the waves to break in the stellar core (e.g. \citealt{2010MNRAS.404.1849B}), though there are other possibilities, including gradual radiative damping of the waves leading to spin-up of the central portions of the star and the subsequent efficient wave absorption (e.g. \citealt{2023MNRAS.521.1353G}), other strong wave-wave interactions (e.g. \citealt{2024ApJ...960...50W}), or sufficiently strong magnetic fields (though this is expected to be more relevant for F-stars; e.g. \citealt{2024ApJ...966L..14D}).

We have computed MESA stellar models and the resulting tidal gravity wave $Q'_\star$ for each of these systems (NGTS-6b and 10b were previously presented in \citealt{2020MNRAS.498.2270B}). For HIP\,65\,A we used a mass $0.781\,\rm M_\odot$ and rotation period of $13.2$ d \citep{Nielsen_2020}, and for WASP-173\,A we used a mass $1.05\,\rm M_\odot$ \citep{2019MNRAS.482.1379H} and rotation period of $8.4$ d \citep{2024A&A...690A.379K} (both with initial metallicity $Z=0.02$). We find $Q'_\star\approx 1.2\times 10^5$ (at age 4.1 Gyr; \citealt{Nielsen_2020}) to be applicable for HIP 65 Ab’s orbital decay, and $Q'_\star\approx 2-6.3\times 10^5$ (at ages 3-7 Gyr, with smaller values for older stars) for WASP-173 Ab. Both NGTS-6b and NGTS-10b are predicted to have $Q'_\star\approx 1\times 10^5$ relevant for their orbital decay at ages of approximately 10 Gyr \citep{2020MNRAS.498.2270B}. These predictions assume the gravity waves to be fully damped, and much larger values ($>10^9$) would be expected if the waves are only weakly damped by radiative diffusion (unless the system happens to excite a g-mode in resonance, which has a low probability). Wave breaking is predicted for the expected ages of WASP-173\,A (i.e. at 3-7~Gyr, with \citealt{2019MNRAS.482.1379H} finding $7\pm3$~Gyr and \citealt{2019ApJS..240...13L} finding $1.5-5.0$~Gyr), so this regime is plausibly justified in this system, though it is less clear in the others, none of which clearly satisfy the criterion for wave breaking (in \S 3.2 of \citealt{2020MNRAS.498.2270B}).

%We test our modified tidal quality factor's lower limit by comparing to theoretical models. Figure 5 from \cite{2024ApJ...960...50W} shows the theoretical order of magnitude predictions of $Q_{\star}'$ for for systems with known stellar mass, stellar age, current orbital period and time averaged decay time. All but the decay time $\langle\tau\rangle$ are known here and appear either in this study or the discovery papers. We calculate $\langle\tau\rangle$ by rewriting Equation 14 from the same study to be \(\langle\tau\rangle = 1.5 P/\dot{P}\).

%It follows that \(\langle\tau\rangle = 10^{9.9\pm0.4} \rm yr\), which corresponds to a theoretical inequality of $Q_{\star}' > 10^{7}$. 

%\textbf{JKT: doesn't the Qstar equation have P in as well as Pdot?}

%\textbf{LG: It shouldn't in the way I've written it here. Orbital period $P$ and quadratic coefficient $p$ are related to $\dot{P}$ through the equation \(p = \frac{1}{2}P\dot{P}\). The equation for $Q_{\star}'$ here is similar to the one in section 7.2 of \cite{2024PSJ.....5..163A}. I'm also pretty sure $Q_{\star}'$ is dimensionless, so dimensional analysis of this equation checks out as well.}

%\textbf{JKT: there definitely is a Porb in there. Check out 2018AcA....68..371M. I happen to have a derivation of this equation from Adrian Barker and that also has a P in, in the form of $2\pi/P$. See also eq.5 in 2017ApJ...836L..24W. I take your point about what the quad coeff $p$ means, but Wilkins+2017 make this clear.}

\section{Results and Discussion}

A transit timing analysis was carried out on four hot Jupiter systems: HIP\,65\,A, NGTS-6, NGTS-10, WASP-173\,A. Their large mass ratios, short periods and large fractional radii made these planets excellent candidates for the detection of tidal decay. We collected a total of 33 transit times of our own from ground-based observations using the 1.54m Danish Telescope, 24 from TESS (two per sector), 21 from the literature and seven from the ETD. The time baselines covered by all data used in this study are $6.4\rm\,yr$, $7\rm\,yr$, $8.7\rm\,yr$ and $9\rm\,yr$ for HIP\,65\,Ab, NGTS-6b, NGTS-10b and WASP-173\,Ab respectively. The results from our analyses are tabulated in \autoref{table stats} and displayed visually in Fig.\,\ref{fig resid}.

\subsection{HIP~65~Ab}

%\textbf{JKT: Maybe think of iterating the published Pdot values here?}

%\textbf{LG: Done. The ones for WASP-173 are in this section now too.}

From the orbital decay model, we find a positive quadratic coefficient, corresponding to a period increase of $7.5 \pm 5.4\rm \,ms\,yr^{-1}$. For comparison, the published period derivatives for HIP\,65\,Ab are $22.3 \pm 9.6 \rm \,ms\,yr^{-1}$ from \cite{2024PSJ.....5..163A} and a rate of period change per cycle of $(-0.1 \pm 3.1)\times10^{-10}\rm\,d$ by \cite{2024A&A...692A..35M}. Our result is consistent with these but also with a constant period at under 2$\sigma$. The statistical analysis shows no substantial favouring of either model, with $\Delta{\rm BIC} = -1.5$ and $\Delta{\rm AIC} = -0.1$, acting as sufficient reason not to explore models with higher-order polynomials. The detection of any orbital decay in this system would require observations over a longer time span than currently available.

%Although we make little reference to the tidal mechanisms within the host star in this work, it may be worth noting the limits of $Q_{\star}'$ from other studies. Theoretical constraints usually vary with additional factors such as stellar age, wave breaking \citep{2010MNRAS.404.1849B} and radiative damping \citep{1975A&A....41..329Z}. For HIP\,65\,A, wave breaking is not predicted for its age and so rapid decay is unlikely \citep{2024A&A...692A..35M}. Our observational constraint for $Q_{\star}'$ and non-detection of orbital decay are consistent with that of \cite{2024PSJ.....5..163A} who found $Q_{\star}'\gtrapprox 10^{5.3}$ and \cite{2024A&A...692A..35M} who found $Q_{\star}'\gtrapprox 10^{4.9}$.

%Pulling the $M_{p}$,$M_{\star}$ and $a/R_{\star}$ from the discovery paper, we find \(Q_{\rm *,max}' \approx (5.08 \pm 0.47) \times 10^{4} \approx 10^{4.71 \pm 0.04}\).

%\textbf{JKT: based on our understanding of tides, the system definitely IS inspiralling (note the double l because we're not American). What you intend to say is that the available data are currently consistent with a constant orbital period.}
%\textbf{LG: I must've misunderstood what the word means. Changed now.}

\subsection{NGTS-6 b}

%\textbf{JKT: how significant are these differences in BIC and AIC?}

%\textbf{LG: Discussed here now and at the end of Section 3.3}

We find the difference between the strength of the linear and quadratic models to be $\Delta{\rm BIC} = 2.3$ and $\Delta{\rm AIC} = 3.5$, which does not significantly favour the quadratic model. Alongside an inspiral rate of $-21 \pm 9\rm \,ms\,yr^{-1}$, tidal decay is not significantly favoured over the constant-period alternative. Although the best-fitting value is negative, this result is consistent with zero at 2.3$\sigma$ and therefore does not directly imply a shrinking orbit. Further observations will be paramount in distinguishing orbital decay from typical orbital motion. Opportunely, NGTS-6\,b is in a sample of $\sim100$ known transiting exoplanets that will be observed by the PLATO mission \citep{Rauer+25exa} in its first long-pointing field (LOPS2; \citealt{2025A&A...694A.313N}). As a result, the occasion may present itself in less than a decade with the release of PLATO data. This planet makes an interesting tidal decay candidate for future study.

%The modified tidal quality factor was derived as \(Q_{\star}' > (1.51 \pm 0.14) \times 10^{4} \approx 10^{4.18 \pm 0.04}\).
%Despite wave breaking being unlikely for the host star, \cite{2020MNRAS.498.2270B} estimates $Q_{\star}'\approx 10^5$ with fully damped gravity waves.

\subsection{NGTS-10 b}

A successful transit timing analysis of NGTS-10 is hindered by the lack of available transit times. Although the times used in this study span over eight years, only 12 transit times have been obtained. The $\dot{P}$ value presented here is negative with a significance of just above 2$\sigma$ based on the available data. There is a gap of several years between the earlier times and the observations presented in the current work (see Fig.\,\ref{fig resid}), meaning that a linear ephemeris is precisely anchored but that the quadratic term is not well constrained.

Without precise points in the epoch space between the original discovery observations and the first Danish Telescope observation, this decay rate could vary significantly in either direction. The status of orbital decay in this system will be helped by new timing data, particularly as this planet is also in the PLATO LOPS2 field \citep{2025A&A...694A.313N} and has planned observations with the JWST \citep{2006SSRv..123..485G,2024AJ....167..195C}. Although theoretical studies provide conflicting evidence for the rate of tidal dissipation (e.g.\ \citealt{2020MNRAS.498.2270B,2024ApJ...973..128T}), longer photometric coverage will be beneficial in constraining $\dot{P}$ even should it be consistent with zero. As of now, NGTS-10\,b still remains a good candidate.

\subsection{WASP-173~Ab}

Our timing analysis of WASP-173\,Ab produced a lower $\dot{P}$ and $p$ than those detected in other works, with \cite{2024PSJ.....5..163A} finding $\dot{P} = 19.3 \pm 11.0 \rm\,ms\,yr^{-1}$ and \cite{2024A&A...692A..35M} finding $p = (0.6 \pm 2.6)\times10^{-10}$. The differences are likely attributed to the variety of data sources used in each work. We find $\Delta{\rm BIC} = 11.5$ and $\Delta{\rm AIC} = 12.5$, indicating the quadratic model fits the residuals better than that of the linear. Although $\dot{P}$ is negative at 3.8$\sigma$, this result could be optimistic, caused by underestimated uncertainties on some of the transit times. The orbital decay scenario is still not conclusive. Nevertheless, \cite{2024A&A...692A..35M} predict wave breaking provided the stellar host is older than $\sim4.5\rm\,Gyr$, giving the theoretical constraint $Q_{\star}'<1.5\times 10^6$ which agrees with our own theoretical estimation. As with the other three systems, more timing data spanning a longer time interval will be needed to identify orbital decay in WASP-173\,A.

\section{Conclusions}

By taking the 3$\sigma$ lower limits of $\dot{P}$ for each system, we computed lower limits for $Q_{\star}'$. These are not remarkable, approximately lying in the range $10^4<Q_{\star,\rm min}'<10^5$, but are compatible with our theoretical predictions if tidally-driven orbital decay due to gravity waves is actually occurring. Given that the current constraints are not very far from predictions, each of these 4 systems would be promising for future follow-up studies since they have very good potential to test tidal theory. On the other hand, the possible outward migration of HIP\,65\,Ab, if confirmed in future studies, would be more difficult to explain by tidal processes alone in such a slowly rotating star. A significant positive $\dot{P}$, if found, would be most likely explained by another phenomenon, such as the light time effect (e.g. \citealt{2025ApJ...986..117Y}) or an acceleration of the system away from the observer. As of now, our TTV models show no conclusive evidence for orbital decay in any of these systems. WASP-173\,Ab has the strongest suggestion of orbital decay and is a promising candidate for future studies.

The data we presented and analysed in the current work will be valuable for future studies of these objects. Long-term monitoring of all four planetary systems will be useful in progressively constraining $Q_{\star}'$ and ultimately leading to a significant detection of tidal effects.

%The most likely cause behind a significant positive $\dot{P}$ would not be tidal growth for this reason. \refa{p1l8: For HIP 65 Ab, the authors interpret the apparent positive period variation as potentially indicative of outward migration and note that such a process would be difficult to reconcile with tidal evolution alone in a slowly rotating host star. While this interpretation is not incorrect, I believe alternative explanations should also be discussed. In particular, they should consider whether the observed trend could be influenced by a light-time effect induced by an additional companion (as recently demonstrated for the HAT-P-7 system by Yang et al. 2025) or by the acceleration of the system during its motion away from the observer. Even if a full analysis is beyond the scope of the paper, mentioning these possibilities (with proper references) would provide a more balanced interpretation of the result.}

%The last numbered section should briefly summarise what has been done, and describe
%the final conclusions which the authors draw from their work.

\section*{Acknowledgements}

JS acknowledges support from STFC under grant number ST/Y002563/1. FA, MA, and UGJ acknowledge funding from the Novo Nordisk Foundation Interdisciplinary Synergy Programme grant no. NNF19OC0057374. VB and PR are supported by PRIN 2022 CUP D53D23002590006. RFJ acknowledges the support provided by the GEMINI/ANID project under grant number 32240028, by ANID’s Millennium Science Initiative through grant ICN12\_009, awarded to the Millennium Institute of Astrophysics (MAS), and by ANID’s Basal project FB210003. EK is supported by the National Research Foundation of Korea 2021M3F7A1082056. LM acknowledges the financial contribution from PRIN MUR 2022 project 2022J4H55R. This paper includes data collected by the TESS\ mission and obtained from the MAST data archive at the Space Telescope Science Institute (STScI). Funding for the TESS\ mission is provided by the NASA's Science Mission Directorate. STScI is operated by the Association of Universities for Research in Astronomy, Inc., under NASA contract NAS 5–26555. This work has made use of data from the European Space Agency (ESA) mission {\it Gaia}\footnote{\texttt{https://www.cosmos.esa.int/gaia}}, processed by the {\it Gaia} Data Processing and Analysis Consortium DPAC\footnote{\texttt{https://www.cosmos.esa.int/web/gaia/dpac/consortium}}). Funding for the DPAC has been provided by national institutions, in particular the institutions participating in the {\it Gaia} Multilateral Agreement. This research has received funding from the Europlanet 2024 Research Infrastructure (RI) programme. The Europlanet 2024 RI provides free access to the world’s largest collection of planetary simulation and analysis facilities, data services and tools, a ground-based observational network and programme of community support activities. Europlanet 2024 RI has received funding from the European Union’s Horizon 2020 research and innovation programme under grant agreement No. 871149.

The following resources were used in the course of this work: the NASA Astrophysics Data System; the SIMBAD database operated at CDS, Strasbourg, France; and the ar$\chi$iv scientific paper preprint service operated by Cornell University.

This work is based on data collected by MiNDSTEp with the Danish 1.54\,m telescope at the ESO La Silla Observatory in Chile. The operation, servicing, and maintenance of the DK-1.54m telescope is supported by a Villum Young Investigator grant (project number 25501) and a Villum Experiment grant (VIL69896) from VILLUM FONDEN.

%%%%%%%%%%%%%%%%%%%%%%%%%%%%%%%%%%%%%%%%%%%%%%%%%%
\section*{Data Availability}

The light curves obtained with the Danish Telescopes will be made available at the Centre de Donn\'ees astronomiques de Strasbourg (CDS) at \texttt{http://cdsweb.u-strasbg.fr/}. The TESS data used in this article are available in the MAST archive (\texttt{https://mast.stsci.edu/portal/Mashup/Clients /Mast/Portal.html}).



\bibliographystyle{mnras}
\bibliography{hip65} % if your bibtex file is called example.bib

%\begin{thebibliography}{99}
%\end{thebibliography}

%%%%%%%%%%%%%%%%%%%%%%%%%%%%%%%%%%%%%%%%%%%%%%%%%%

%%%%%%%%%%%%%%%%% APPENDICES %%%%%%%%%%%%%%%%%%%%%

\appendix

\section{Transit light curves from TESS}

\begin{figure*}
    \centering
    \includegraphics[width=1\linewidth]{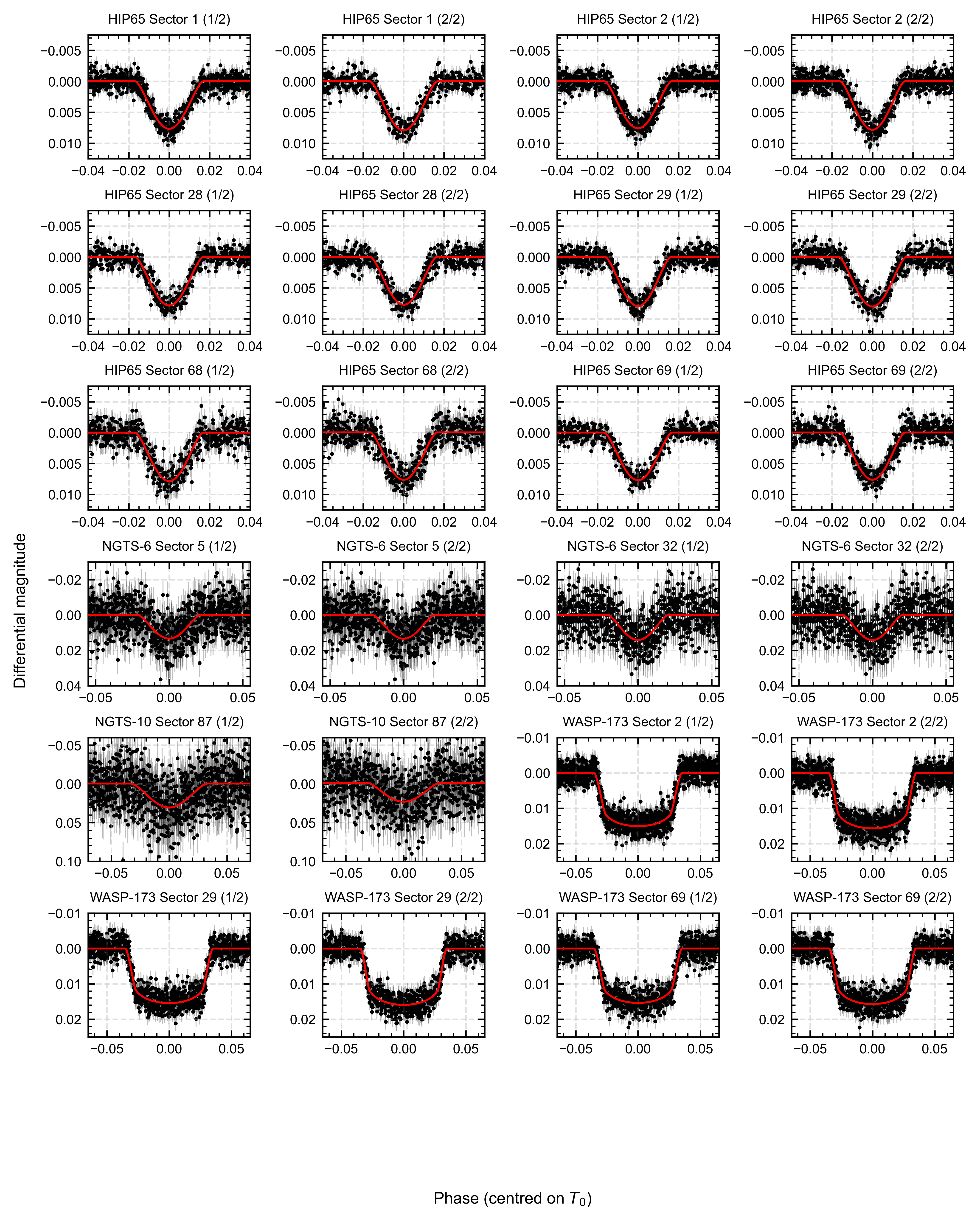}
    \caption{Phase-folded transit light curves for HIP\,65\,A, NGTS-6, NGTS-10 and WASP-173\,A taken from TESS. Data points are displayed in black with their errorbars. Fits are plotted in red. The TESS sectors are displayed above each plot. 1/2 refers to the first half of a sector. 2/2 refers to the second half of a sector.}
    \label{fig:tess}
\end{figure*}

\section{Timing data used in this work}

\begin{table*}
    \centering
    \setlength{\tabcolsep}{30pt}
    \caption{Times of mid-transit for HIP\,65\,A used in this study.}
    \begin{filecontents*}{TableTTs.csv}
BJD(TDB),Epoch,Residuals (d),Source
2458331.98999 ± 0.00012,-1232.0,-0.00004,This work (TESS)
2458343.76152 ± 0.00015,-1220.0,-0.00018,This work (TESS)
2458360.43826 ± 0.00011,-1203.0,+0.00004,This work (TESS)
2458375.15301 ± 0.00012,-1188.0,+0.00020,This work (TESS)
2459066.73824 ± 0.00019,-483.0,+0.00004,This work (TESS)
2459079.49091 ± 0.00015,-470.0,+0.00007,This work (TESS)
2459094.20536 ± 0.00015,-455.0,-0.00006,This work (TESS)
2459107.93893 ± 0.00013,-441.0,-0.00010,This work (TESS)
2459477.76576 ± 0.00022,-64.0,+0.00022,This work (Danish Telescope)
2459479.72753 ± 0.00026,-62.0,+0.00004,This work (Danish Telescope)
2459538.58580 ± 0.00080,-2.0,-0.00002,Kokori et al. 2023
2459539.56570 ± 0.00110,-1.0,-0.00109,Kokori et al. 2023
2459540.54710 ± 0.00070,0.0,-0.00066,Kokori et al. 2023
2459837.78214 ± 0.00031,303.0,-0.00020,This work (Danish Telescope)
2459838.76421 ± 0.00044,304.0,+0.00090,This work (Danish Telescope)
2459843.66721 ± 0.00023,309.0,-0.00096,This work (Danish Telescope)
2460140.90250 ± 0.00018,612.0,-0.00024,This work (Danish Telescope)
2460142.86476 ± 0.00023,614.0,+0.00007,This work (Danish Telescope)
2460159.54129 ± 0.00022,631.0,+0.00008,This work (TESS)
2460173.27477 ± 0.00022,645.0,-0.00005,This work (TESS)
2460187.00852 ± 0.00022,659.0,+0.00009,This work (TESS)
2460199.76095 ± 0.00025,672.0,-0.00012,This work (TESS)
2460200.74242 ± 0.00023,673.0,+0.00038,This work (Danish Telescope)
2460250.76994 ± 0.00090,724.0,-0.00169,ETD
2460509.74882 ± 0.00021,988.0,+0.00054,This work (Danish Telescope)
2460510.72940 ± 0.00038,989.0,+0.00014,This work (Danish Telescope)
2460557.81569 ± 0.00032,1037.0,-0.00023,This work (Danish Telescope)
2460566.64733 ± 0.00154,1046.0,+0.00266,ETD
2460674.55074 ± 0.00103,1156.0,-0.00087,ETD
\end{filecontents*}
\csvautotabular{TableTTs.csv}
    \label{table time}
\end{table*}

\begin{table*}
    \centering
    \setlength{\tabcolsep}{30pt}
    \caption{Times of mid-transit for NGTS-6 used in this study.}
    \begin{filecontents*}{TableTTs_ngts6_.csv}
BJD(TDB),Epoch,Residuals (d),Source
2457982.37942 ± 0.00040,-1778.0,-0.00056,This work (Vines et al. 2019)
2458443.69673 ± 0.00056,-1255.0,+0.00033,This work (TESS)
2458457.80946 ± 0.00053,-1239.0,+0.00013,This work (TESS)
2458718.89791 ± 0.00047,-943.0,-0.00064,This work (Danish Telescope)
2458733.89478 ± 0.00044,-926.0,+0.00124,This work (Danish Telescope)
2458741.83130 ± 0.00041,-917.0,-0.00076,This work (Danish Telescope)
2459181.09727 ± 0.00070,-419.0,+0.00025,This work (TESS)
2459194.32695 ± 0.00058,-404.0,-0.00094,This work (TESS)
2459482.76136 ± 0.00037,-77.0,+0.00045,This work (Danish Telescope)
2459489.81758 ± 0.00038,-69.0,+0.00020,This work (Danish Telescope)
2459549.79580 ± 0.00230,-1.0,-0.00153,Kokori et al. 2023
2459549.79920 ± 0.00120,-1.0,+0.00187,Kokori et al. 2023
2459550.68140 ± 0.00080,0.0,+0.00201,Kokori et al. 2023
2459550.68360 ± 0.00180,0.0,+0.00421,Kokori et al. 2023
2459557.73320 ± 0.00070,8.0,-0.00265,Kokori et al. 2023
2459558.61710 ± 0.00090,9.0,-0.00081,Kokori et al. 2023
2459558.61910 ± 0.00100,9.0,+0.00119,Kokori et al. 2023
2459573.61300 ± 0.00100,26.0,+0.00010,Kokori et al. 2023
2459863.81018 ± 0.00025,355.0,+0.00015,This work (Danish Telescope)
2459953.78172 ± 0.00054,457.0,+0.00175,ETD
2460222.80771 ± 0.00033,762.0,+0.00001,This work (Danish Telescope)
2460230.74805 ± 0.00391,771.0,+0.00183,ETD
2460558.87158 ± 0.00024,1143.0,-0.00028,This work (Danish Telescope)
2460566.80843 ± 0.00162,1152.0,-0.00196,ETD
2460566.81014 ± 0.00030,1152.0,-0.00024,This work (Danish Telescope)
\end{filecontents*}
\csvautotabular{TableTTs_ngts6_.csv}
    \label{table time2}
\end{table*}

\begin{table*}
    \centering
    \setlength{\tabcolsep}{30pt}
    \caption{Times of mid-transit for NGTS-10 used in this study.}
    \begin{filecontents*}{TableTTs_ngts10_.csv}
BJD(TDB),Epoch,Residuals,Source
2457518.84457 ± 0.00021,-3282.0,-0.00019,This work (McCormac et al. 2020)
2458038.79843 ± 0.00020,-2604.0,+0.00001,This work (McCormac et al. 2020)
2458085.57890 ± 0.00063,-2543.0,-0.00002,This work (McCormac et al. 2020)
2458104.75168 ± 0.00043,-2518.0,+0.00043,This work (McCormac et al. 2020)
2459834.86286 ± 0.00024,-262.0,+0.00028,This work (Danish Telescope)
2459844.83251 ± 0.00025,-249.0,+0.00032,This work (Danish Telescope)
2460226.74450 ± 0.00022,249.0,-0.00056,This work (Danish Telescope)
2460229.81299 ± 0.00019,253.0,+0.00036,This work (Danish Telescope)
2460571.84743 ± 0.00059,699.0,+0.00038,This work (Danish Telescope)
2460581.81646 ± 0.00017,712.0,-0.00020,This work (Danish Telescope)
2460672.30953 ± 0.00091,830.0,-0.00054,This work (TESS)
2460686.11444 ± 0.00092,848.0,+0.00028,This work (TESS)
\end{filecontents*}
\csvautotabular{TableTTs_ngts10_.csv}
    \label{table time3}
\end{table*}

\begin{table*}
    \centering
    \setlength{\tabcolsep}{30pt}
    \caption{Times of mid-transit for WASP-173\,A used in this study.}
    \begin{filecontents*}{TableTTs_wasp173_.csv}
BJD(TDB),Epoch,Residuals,Source
2457261.12650 ± 0.00130,-1605.0,+0.00030,Labadie-Bartz et al. 2019
2457682.66826 ± 0.00020,-1301.0,-0.00050,This work (Hellier et al. 2019)
2458048.74549 ± 0.00082,-1037.0,+0.00030,Labadie-Bartz et al. 2019
2458105.59827 ± 0.00088,-996.0,+0.00030,Labadie-Bartz et al. 2019
2458360.74232 ± 0.00024,-812.0,+0.00017,This work (TESS)
2458374.60885 ± 0.00025,-802.0,+0.00017,This work (TESS)
2459092.89522 ± 0.00022,-284.0,+0.00021,This work (TESS)
2459106.76190 ± 0.00019,-274.0,+0.00036,This work (TESS)
2459386.86650 ± 0.00080,-72.0,+0.00102,Kokori et al. 2023
2459486.70408 ± 0.00085,0.0,-0.00043,This work (Danish Telescope)
2459766.80873 ± 0.00025,202.0,+0.00029,This work (Danish Telescope)
2459827.82086 ± 0.00028,246.0,-0.00032,This work (Danish Telescope)
2460157.84486 ± 0.00024,484.0,+0.00023,This work (Danish Telescope)
2460182.80317 ± 0.00031,502.0,-0.00121,This work (Danish Telescope)
2460186.96458 ± 0.00036,505.0,+0.00024,This work (TESS)
2460200.83102 ± 0.00020,515.0,+0.00014,This work (TESS)
2460469.83711 ± 0.00114,709.0,-0.00447,ETD
2460494.80195 ± 0.00048,727.0,+0.00061,This work (Danish Telescope)
2460544.71962 ± 0.00046,763.0,-0.00124,This work (Danish Telescope)
\end{filecontents*}
\csvautotabular{TableTTs_wasp173_.csv}
    \label{table time4}
\end{table*}

%\section{TESS transit light curves}


%\begin{figure*}
%    \centering
%    \includegraphics[width=1\linewidth]{TessFig3.png}
%    \caption{Phase folded transit light curves for the HIP\,65\,A system taken by TESS. Data points are shown in blue with their errorbars. Fits are overplotted in red. The TESS sector is displayed above each plot.}
%    \label{tessfig}
%\end{figure*}

%\begin{figure*}
%    \centering
%    \includegraphics[width=1\linewidth]{TessFig3(wasp173).png}
%    \caption{Phase folded transit light curves for the WASP-173\,A system taken by TESS. Data points are shown in blue with their errorbars. Fits are overplotted in red. The TESS sector is displayed above each plot.}
%    \label{tessfig2}
%\end{figure*}

%\begin{figure*}
%    \centering
%    \includegraphics[width=1\linewidth]{TessFig3(ngts6).png}
%    \caption{Phase folded transit light curves for the NGTS-6 system taken by TESS. Data points are shown in blue with their errorbars. Fits are overplotted in red. The TESS sector is displayed above each plot.}
%    \label{tessfig3}
%\end{figure*}

%%%%%%%%%%%%%%%%%%%%%%%%%%%%%%%%%%%%%%%%%%%%%%%%%%

% Don't change these lines
\bsp	% typesetting comment
\label{lastpage}
\end{document}